\documentclass[aps,prx,reprint,preprintnumbers,superscriptaddress,nofootinbib,longbibliography,floatfix]{revtex4-2}
\pdfoutput=1
\usepackage{rotating}
\usepackage{array}
\usepackage{amsmath}
\usepackage[normalem]{ulem}
\usepackage{slashed}
\usepackage{booktabs}
\usepackage[pdftex,table]{xcolor}
\usepackage{units}
\usepackage{xfrac}
\usepackage{mathtools}
\usepackage{empheq}
\usepackage[]{units}
\usepackage{multirow}
\usepackage{amssymb}
\usepackage{url}
\usepackage{comment}
\usepackage{physics}
\usepackage{color,soul}
\usepackage{bbm}
\usepackage[caption=false]{subfig}
\usepackage{adjustbox}
\usepackage[T1]{fontenc}

\usepackage{hyperref}
\hypersetup{
  colorlinks=true,
  citecolor=blue,
  linkcolor=blue,
  urlcolor=blue
}

\begin{document}

\title{The Fundamental Limit of Jet Tagging}

\author{Joep Geuskens}
\email{joep.geuskens@rwth-aachen.de}
\affiliation{Institute for Theoretical Particle Physics and Cosmology, RWTH Aachen University, Germany}

\author{Nishank Gite}
\email{nishankgite@berkeley.edu}
\affiliation{Department of Physics, University of California, Berkeley, CA 94720, USA}

\author{Michael Kr\"amer}
\email{mkraemer@physik.rwth-aachen.de}
\affiliation{Institute for Theoretical Particle Physics and Cosmology, RWTH Aachen University, Germany}

\author{Vinicius Mikuni}
\email{vmikuni@lbl.gov}
\affiliation{National Energy Research Scientific Computing Center, Berkeley Lab, Berkeley, CA 94720, USA}

\author{Alexander M\"uck}
\email{mueck@physik.rwth-aachen.de}
\affiliation{Institute for Theoretical Particle Physics and Cosmology, RWTH Aachen University, Germany}

\author{Benjamin Nachman}
\email{bpnachman@lbl.gov}
\affiliation{Physics Division, Lawrence Berkeley National Laboratory, Berkeley, CA 94720, USA}
\affiliation{Berkeley Institute for Data Science, University of California, Berkeley, CA 94720, USA}

\author{Humberto Reyes-Gonz\'alez}
\email{humberto.reyes@rwth-aachen.de}
\affiliation{Institute for Theoretical Particle Physics and Cosmology, RWTH Aachen University, Germany}

\begin{abstract}
    Identifying the origin of high-energy hadronic jets (`jet tagging') has been a critical benchmark problem for machine learning in particle physics.  Jets are ubiquitous at colliders and are complex objects that serve as prototypical examples of collections of particles to be categorized.  Over the last decade, machine learning-based classifiers have replaced classical observables as the state of the art in jet tagging.  Increasingly complex machine learning models are leading to increasingly more effective tagger performance.  Our goal is to address the question of convergence - are we getting close to the fundamental limit on jet tagging or is there still potential for computational, statistical, and physical insights for further improvements?  We address this question using state-of-the-art generative models to create a realistic, synthetic dataset with a known jet tagging optimum.  Various state-of-the-art taggers are deployed on this dataset, showing that there is a significant gap between their performance and the optimum.  Our dataset and software are made public to provide a benchmark task for future developments in jet tagging and other areas of particle physics.
\end{abstract}

\maketitle

Identifying the origin of hadronic jets has been a central task in collider physics for decades.  A single high-energy jet at the Large Hadron Collider can be composed of hundreds of particles, each with a four-vector and other properties.  This makes jet classification (`tagging') a high-dimensional inference task.  Until a few years ago, most of the progress in jet tagging was made through the development of physics-informed observables based on the radiation pattern within jets (`jet substructure')~\cite{Larkoski:2017jix,Kogler:2018hem}.  Modern machine learning has led to rapid progress in jet tagging, leading to many factors of improvement over traditional algorithms (see e.g. Ref.~\cite{Kasieczka:2019dbj,CMS:2020poo,Qu:2019gqs,Qu:2022mxj,ATLAS:2022qby}). There is seemingly no end to the performance gains.  

In the language of statistical learning, we know that there is a theoretical upper bound to the sensitivity of a jet tagging algorithm: \textit{the likelihood ratio}~\cite{Neyman289}.  Given two classes of jets $A$ and $B$ and a representation of a jet $x\in\mathbb{R}^N$, any monotonic function of $p(x|A)/p(x|B)$ is the optimal classifier - there is no other observable that can have a lower probability of miss-tagging a $B$ jet as an $A$ jet (false positive rate) for a given probability of correctly identifying an $A$ jet as an $A$ jet (true positive rate).  In practice, we do not have direct access to the likelihoods $p(x|A)$ and $p(x|B)$, especially since $x$ can be high dimensional and jets have a stochastic number of constituents; jets are thus \textit{point clouds}. For this reason, we do not know how close existing approaches are to the theoretical optimum.

Modern machine learning has given us another tool that may help us answer this question: generative AI.  Neural networks are capable of approximating $p(x|A)$ and $p(x|B)$ and can also sample from the probability densities.  Generative networks have been studied for many years now in collider physics (see e.g. Ref.~\cite{deOliveira:2017pjk,Butter:2022rso,Adelmann:2022ozp}), but it is only recently that we have models that both can parse the structure of jets as full (variable-length) point clouds and estimate the probability density at the same time.  Such methods include normalizing flows~\cite{Verheyen:2022tov,Kach:2022qnf,Kach:2022uzq,Schnake:2024mip}, diffusion models~\cite{Mikuni:2023tok}, and auto-regressive models~\cite{Finke:2023veq}.  In this paper, we use the auto-regressive model of Ref.~\cite{Finke:2023veq} to create a highly realistic, synthetic dataset where we know the per-jet likelihood ratio.  With this dataset, we train state-of-the-art classifiers to determine how close existing methods are to the theoretical optimum.  The dataset and code are published alongside this paper, to serve as a benchmark task for future developments in jet tagging and in other areas of particle physics.

As a numerical example, we use the  JetClass dataset from Ref.~\cite{Qu:2022mxj,JetClass} to train our generative model. Top quarks almost always decay to a $b$ quark and a $W$ boson~\cite{Tanabashi:2018oca}. When the top quark is highly Lorentz-boosted in the lab frame, its decay products are collimated and collected within a single jet.  In the all-hadronic channel, the $W$ boson decays into quarks and so top quark jets tend to have 3-prong substructure.  In contrast, generic quark and gluon jets typically have much less structure.  As boosted top quarks are a signature of many new physics scenarios, they serve as an important test bed for new methods.  In the JetClass dataset, each jet is represented by its constituents: reconstructed particles.  These jets are generated with MadGraph5\_aMC@NLO~\cite{Alwall:2014hca}, and Pythia~8~\cite{Sjostrand:2006za,Sjostrand:2014zea} is used for showering and hadronization. Delphes~3~\cite{deFavereau:2013fsa,Mertens:2015kba,Selvaggi:2014mya} and the CMS detector card provides the detector simulation and reconstruction based on FastJet~\cite{Cacciari:2011ma}. The dataset includes the 4-momenta of all constituents within each jet. Note that the number of constituents can vary significantly between jets, with an average of about 30 to 50, depending on the jet type.  There are 10M boosted top quark jets and 10M generic quark and gluon (henceforth, QCD) jets.

We train a generative neural network to mimic the QCD and top quark JetClass datasets.  We initially explored the diffusion model from Ref.~\cite{Mikuni:2023tok}, but the probability density is only known approximately.  Current methods for estimating the density do not have the precision required to guarantee that the likelihood ratio is the optimal classifier.  Instead, we employ the autoregressive model of Ref.~\cite{Finke:2023veq}, which provides the exact probability density of each jet.  In this model, the momenta of particles are discretized, ordered in transverse momentum, and then the setup is cast as a natural language problem.  The phase space is finite and each jet can be described by its probability, which is itself a product over conditional probabilities from the constituents parameterized by a transformer~\cite{DBLP:journals/corr/VaswaniSPUJGKP17}.  Generating jets proceeds by sampling from these conditional probabilities in serial.  Previous work has shown that discretization effects are negligible for the classification task, and the resulting model is an accurate representation of the discretized training dataset. While there will be some differences between the surrogate model and the physics model, we believe that these effects will not qualitatively change the conclusions. The physics model itself is only an approximation to real jet data -- we are most concerned that the surrogate model has a similar level of complexity and expressivity as the real model.\footnote{We have trained the autoregressive transformer model of Ref.~\cite{Finke:2023veq} on 10M QCD and 10M top quark jets from the JetClass dataset and verified that a classifier cannot easily distinguish jets generated by our model from the discretized training jets. The extended training data increases the similarity between the generated and training datasets, resulting in even worse classifier performance compared to the results in Ref.~\cite{Finke:2023veq}.}

With this new synthetic dataset, we can now directly compare the performance of various taggers to the optimal performance given by the likelihood ratio. To compare classifier performances we use receiver-operating characteristic (ROC) curves.  These curves show the tradeoff between signal (top tagging) efficiency and background rejection (inverse QCD jet efficiency).  One tagger is uniformly better than another if all of its working points are up and to the right in the plot. 

\begin{figure}[h]
\center
\includegraphics[width=0.475\textwidth]{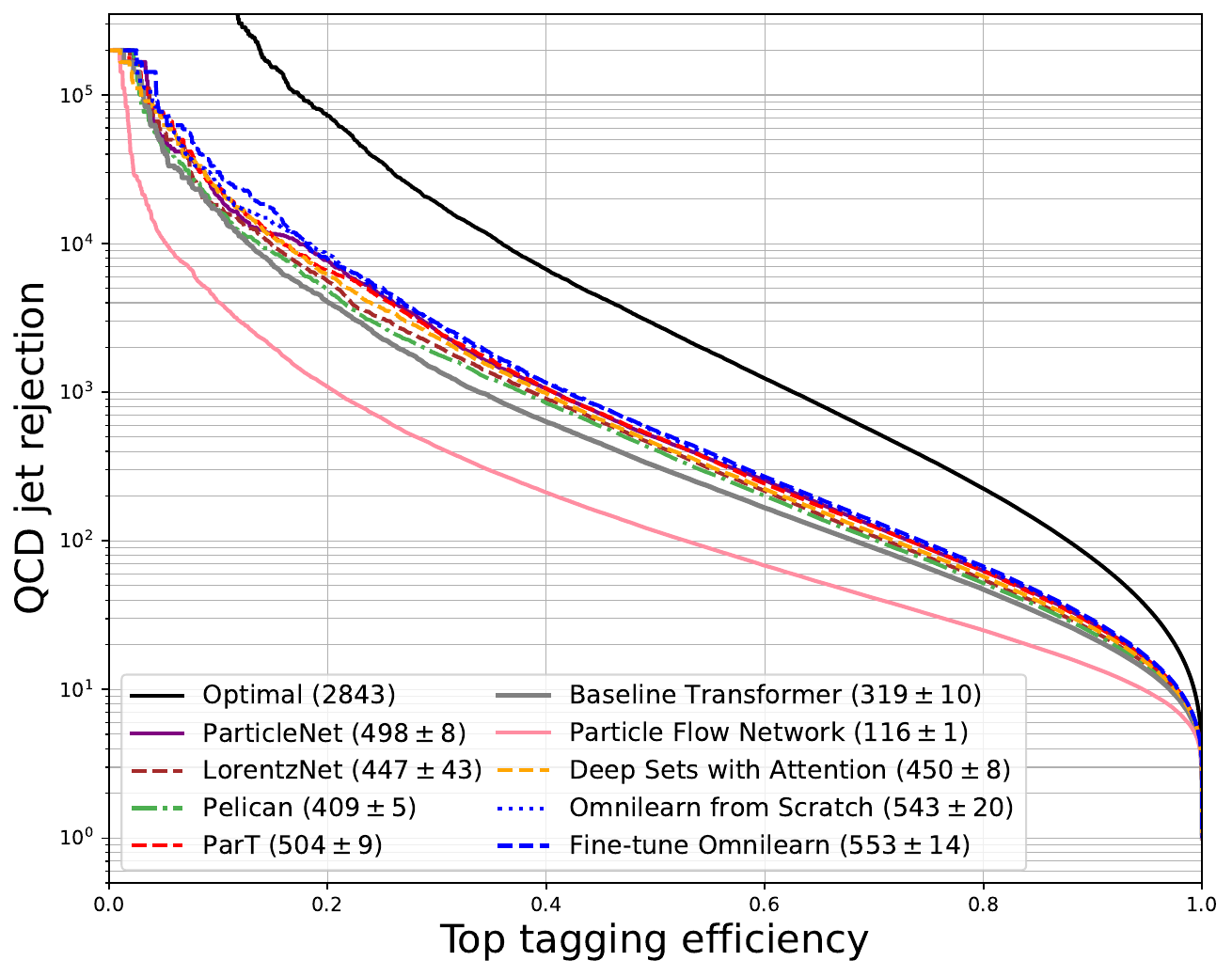}
\caption{\label{fig:compare} ROC curves for top quark versus QCD jet classification.  The performance of the likelihood ratio classifier is labeled 'Optimal' and the other curves correspond to different classifiers trained with 10M jets of each class. The parentheses in the legend indicate the rejection (inverse false positive rate) at a top quark efficiency of 50\% and the corresponding error from training 5 classifiers independently.}
\end{figure}

We fix the size of the classifier training dataset to 10M boosted top quark jets and 10M QCD jets and compare different neural network architectures, similar to the widely-used benchmark comparison of Ref.~\cite{Kasieczka:2019dbj}, only now we know the optimal solution. Since that original work, there have been many innovations in jet tagging and we consider a selection of state-of-the-art and/or widely used models including ParticleNet~\cite{Qu:2019gqs}, Deep Sets~\cite{DBLP:conf/nips/ZaheerKRPSS17} 
%with infrared and collinear (IRC) safety (Energy Flow Networks~\cite{Komiske:2018cqr}), 
without IRC safety (Particle Flow Networks~\cite{Komiske:2018cqr}) and with attention (Deep Sets with Attention~\cite{DBLP:journals/corr/VaswaniSPUJGKP17}); Lorentz-equivariant models such as LorentzNet~\cite{Gong:2022lye} and Pelican~\cite{Bogatskiy:2022czk}; transformers such as ParT~\cite{Qu:2022mxj} 
and OmniLearn (trained from scratch)~\cite{Mikuni:2024qsr} and a classifier based on the transformer architecture of Ref.~\cite{Finke:2023veq}, where we replace the output layer with a binary cross entropy layer (Baseline Transformer); and the fine-tuned OmniLearn foundation model~\cite{Mikuni:2024qsr}. In all cases, we have used the settings described in the corresponding papers without additional hyperparameter optimization.  Our results are displayed in Fig.~\ref{fig:compare}. 
All taggers show comparable performance, which is also similar to the performance achieved on the datasets of Refs.~\cite{Kasieczka:2019dbj,Qu:2022mxj}.  However, even the most powerful current taggers fall well short of optimal performance; for example, with a top-tagging efficiency of 0.5, the taggers exhibit a background rejection of at most about 550, whereas optimal performance corresponds to a background rejection of about 3300. 

\begin{figure}[t]
\center
\includegraphics[width=0.475\textwidth]{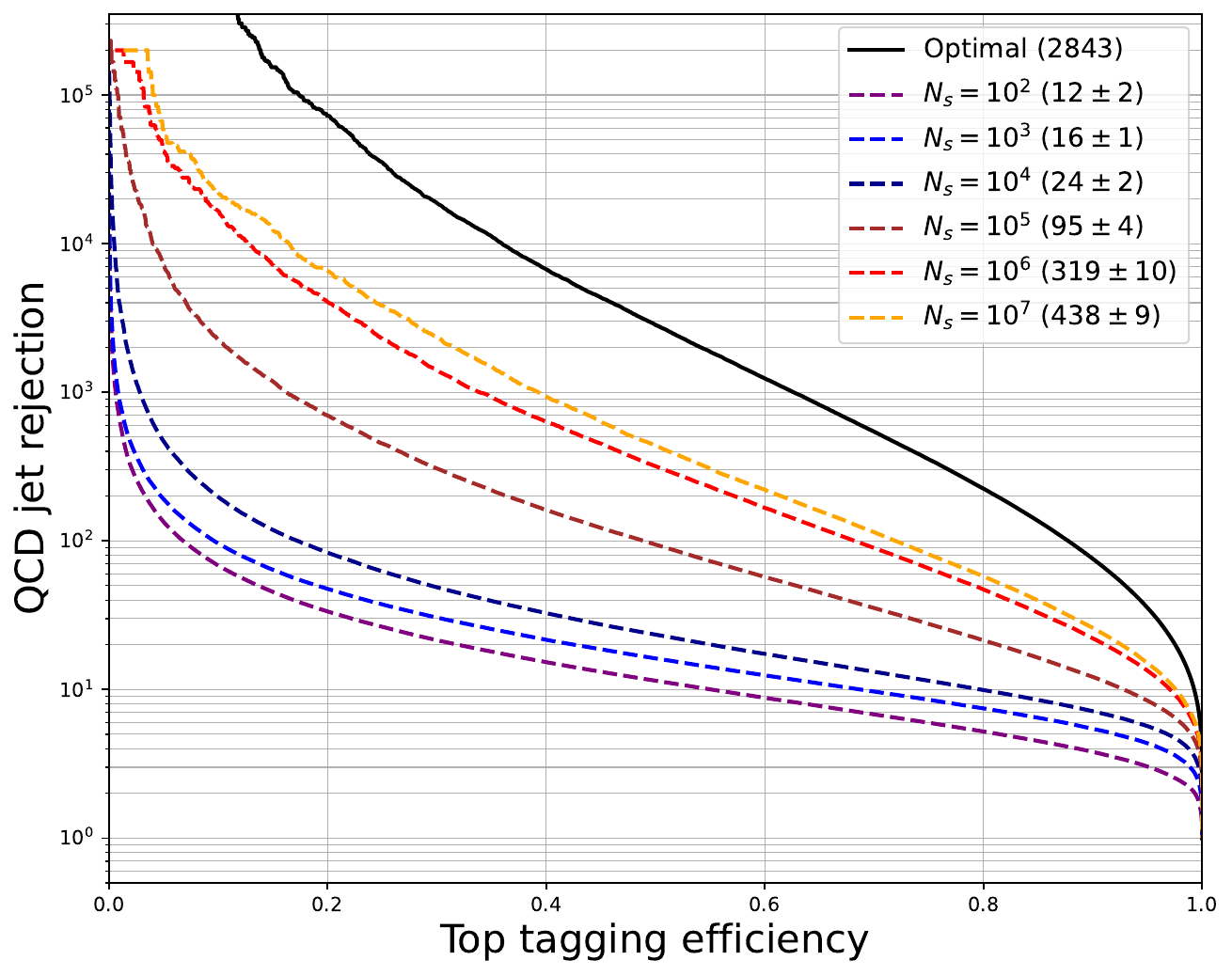}
\caption{\label{fig:d_dep} ROC curves for top quark versus QCD jet classification.  The performance of the likelihood ratio classifier is labelled 'Optimal' and the other curves correspond to the baseline transformer classifier trained on datasets with $N_s$ training jets for each class. The parentheses in the legend indicate the rejection (inverse false positive rate) at a top quark efficiency of 50\% and the corresponding error from training 5 classifiers independently.}
\end{figure}

To analyze our results in more depth, we first investigate how the performance of the taggers varies with the size of the training dataset. To do this, we compare the optimal performance with the ROC curves for a selected tagger trained on training datasets of different sizes, from 100 to 10M events. The results are shown in Fig.~\ref{fig:d_dep} for the classifier based on the transformer architecture of Ref.~\cite{Finke:2023veq} (Baseline Transformer in Fig.~\ref{fig:compare}). Of course, the performance of the tagger initially improves with the size of the training dataset, but the improvement from 1M to 10M training events is rather limited. This means that even with significantly larger training data sets, optimal performance cannot be expected (for fixed architecture). 

\begin{figure}[t]
\center
\includegraphics[width=0.475\textwidth]{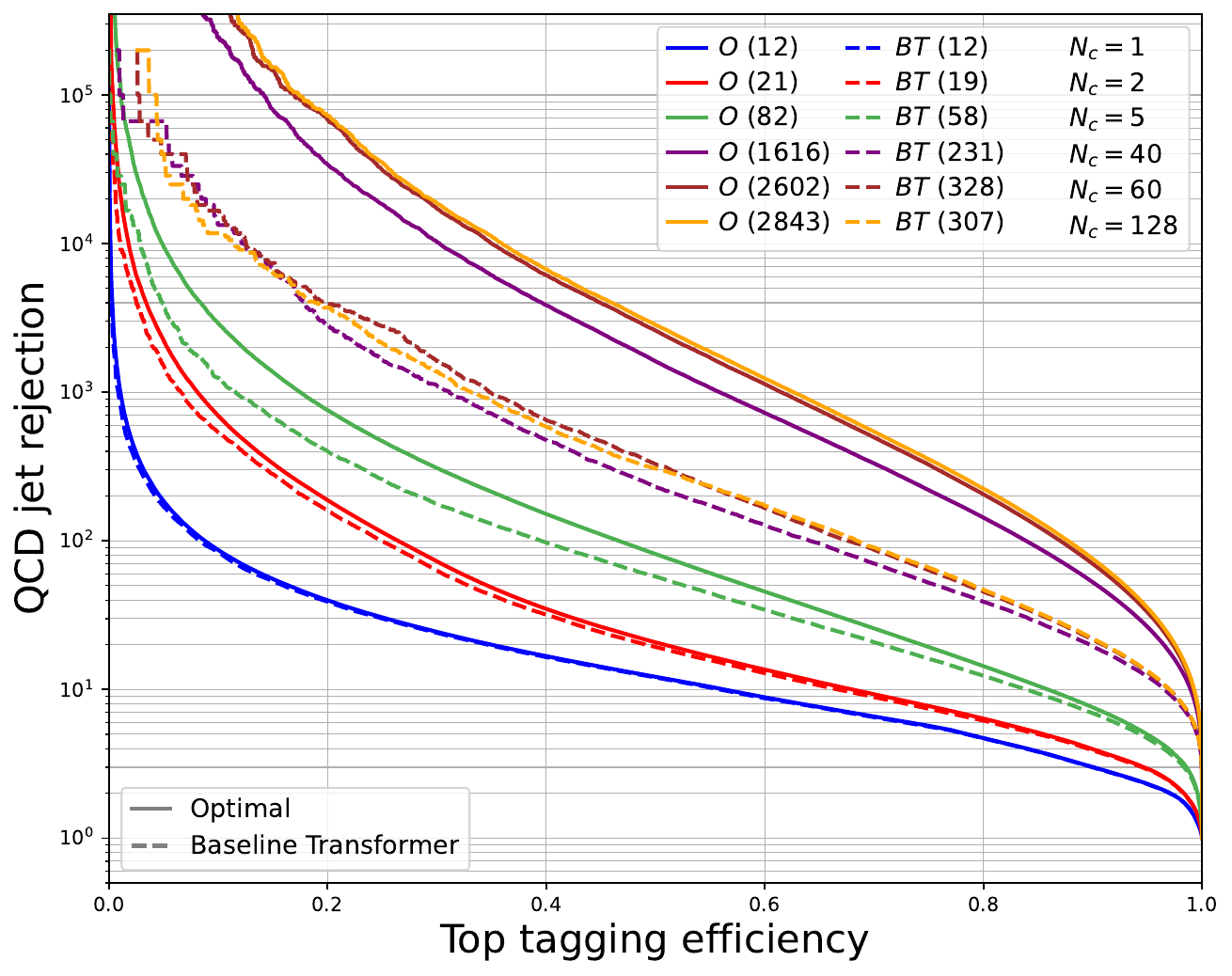}
\caption{\label{fig:nconst} ROC curves for top quark versus QCD jet classification for decreasing number of jet constituents $N_c$.  The performance of the likelihood ratio classifier is shown as solid lines, while the dashed lines correspond to the baseline transformer classifier. The parentheses in the legend indicate the rejection (inverse false positive rate) at a top quark efficiency of 50\%. Note that the dashed lines in this figure correspond to a baseline transformer trained on 1M events.}
\end{figure}

Our results show that even the most advanced top tagger fails to capture the complexity of jets required for optimal classification performance. To investigate this further, we compare the performance of the Baseline Transformer tagger with optimal performance for simpler jets by analyzing only the first $N_c$ constituents as generated by our autoregressive transformer. For $N_c=1$ and $N_c=2$, the low complexity allows any classifier to achieve optimal performance, as shown by the ROC curve in Figure~\ref{fig:nconst}. As $N_c$ increases, jets are classified more efficiently due to more information about their structure. However, when $N_c$ exceeds 5, the classifier performance starts to fall below the optimal performance, suggesting that not all relevant information from additional particles can be extracted. Although specific performance may vary, this trend is expected to hold for any current classifier. Beyond about $N_c= 60$, the optimal performance plateaus, suggesting that additional particles lack information specific to QCD or top jets. This limitation may be due to the limited ability of the generative model to learn jet type specific correlations involving very soft constituents.

In conclusion, we investigated whether state-of-the-art jet taggers are close to optimal or whether there is room for further improvement. Using modern generative models, we created a realistic synthetic dataset for top and QCD jets with known optimal tagging performance. By applying several advanced taggers to this dataset, we found a significant gap between their performance and the optimum. This gap cannot be closed by training on larger datasets and persists because the taggers do not fully capture the complexity of the jets. 

We have made our dataset and software publicly available to serve as a benchmark for future developments in jet tagging and other areas of particle physics.

\section*{Code Availability}
The data for this paper can be found at \url{https://doi.org/10.5281/zenodo.14023638} and code for this paper can be found at \url{https://github.com/hreyes91/fun-jet-tagging}.

\section*{Acknowledgments}
We would like to thank Thorben Finke for his contribution in the early stages of this work. BN thanks Lester Mackey, Ariel Schwartzman, and the Stanford Data Science Initiative (circa 2014) where we asked the same question as in this paper, but without the right tools to provide a sharp answer. NG, VM, and BN are supported by the U.S. Department of Energy (DOE), Office of Science under contract DE-AC02-05CH11231. The research of MK, AM and HR-G is supported by the German Research Foundation DFG under grant 396021762 -- TRR 257: Particle physics phenomenology after the Higgs discovery. This research used resources of the National Energy Research Scientific Computing Center, a DOE Office of Science User Facility supported by the Office of Science of the U.S. Department of Energy under Contract No. DE-AC02-05CH11231 using NERSC awards HEP-ERCAP0021099 and HEP-ERCAP0028249, and computing resources provided by RWTH Aachen University under the project rwth0934. This work was performed in part at the Aspen Center for Physics, supported by National Science Foundation grant PHY-2210452. 

\appendix

\bibliography{HEPML,other}
\bibliographystyle{apsrev4-1}

\clearpage

\end{document}